\documentclass[manuscript]{aastex}

\begin{document}
\pagenumbering{arabic}

\title{SPIRAL STRUCTURE AND GALAXY ENVIRONMENT}

\author{Sidney van den Bergh}
\affil{Dominion Astrophysical Observatory, Herzberg Institute of Astrophysics, National Research Council of Canada, 5071 West Saanich Road, Victoria, BC, V9E 2E7, Canada}
\email{sidney.vandenbergh@nrc.ca}

\begin{abstract}

Among 330 normal spirals of types Sa-Sc the fraction of objects
exhibiting ``ring'', ``intermediate'' and ``spiral'' arm varieties does 
not correlated with environment. A similar conclusion appears to
apply to the arm varieties of 123 barred spirals of types SBa-SBc.
It is concluded that, among the northern Shapley-Ames galaxies, the 
distinction between the spiral and ring varieties of spiral arms is, 
within the accuracy of presently available data, independent of galaxy 
environment. This result suggests that the detailed morphology of
spiral arms depends primarily on parent galaxy characteristics, rather
than on the galactic environment. 

\end{abstract}


\section{INTRODUCTION}

De Vaucouleurs (1959) described the morphology of spiral arms
by dividing them into ring {\it r}, intermediate {\it rs} and spiral {\it s} varieties. Following in de Vaucouleurs's footsteps Sandage \& Tammann 
(1981) used large-scale plates to assign spiral galaxies in the 
Shapley-Ames Catalog (Shapley \& Ames 1931) to the {\it r}, {\it rs}, and {\it s} varieties. Sandage \& Tammann rarely assigned spiral arm varieties 
to galaxies of type Sd, and never to Sm, E/Sa, and S0/a galaxies. 
Furthermore it was only possible for Sandage \& Tammann to assign 
about half of the tightly wound spiral structure in Sa and SBa 
galaxies to the {\it s}, {\it rs}, and {\it r} varieties. In galaxies that were viewed edge-on, or almost edge-on, it was usually not possible to 
determine the arm varieties.  

A book length review on all aspects of galactic rings has been published by Buta \& Combes (1996).  Furthermore Buta (1995) has given a catalog of 3692 ringed galaxies located south of $\delta$ = -17$^{\circ}$.  Finally physical processes that lead to the formation of such rings have been discussed by Buta (1999).

It is the purpose of the present investigation to see if the 
dichotomy between spiral and ring-shaped arms might correlate with 
galaxy environment. To check on this point the environmental 
assignments of Shapley-Ames galaxies north of $\delta$ = -27$^{\circ}$ , that were recently published by van den Bergh (2002), were compared with the
spiral arm varieties listed by Sandage \& Tammann. Galaxies were 
assigned to the ``field'', to ``groups'' or to ``clusters'' on the basis 
of an inspection of their environments on the prints of the Palomar 
Sky Survey and/or their group or cluster designations in van den 
Bergh (1960). A galaxy was regarded as a group member if it appeared 
to belong to a grouping that had between three and six other (non-dwarf) 
members. A galaxy had to have at least six (non-dwarf) companions to 
be assigned to the cluster category. More details on how environmental 
assignments were made are given in van den Bergh (2002). In that 
paper it was found that the frequency distribution of Hubble types
along the sequence S0-Sa-Sb-Sc-Sm-Im in ``field'' areas differed from
the morphological frequency distribution in ``cluster'' areas at greater 
than 99.9\% confidence. The observed differences were in the expected 
sense (Hubble 1936, p.79, Dressler 1980), with early-type galaxies being
over abundant in cluster areas and under abundant in the field. This 
result strengthens confidence in the conclusion that the ``field'',
``group'', and "cluster" assignments correspond to significant physical differences in galaxy environment. 
 
\section{FIELD GALAXIES}

Since field galaxies are expected to suffer fewer encounters with 
neighbors, than do members of groups and clusters, they provide a 
fiducial against which group and cluster members can be compared. For 
172 pure spirals of types Sa-Sc, 69\% are of the {\it s} variety, 10\% are of type {\it rs} and 21\% are of the {\it r} variety. For 61 barred spirals of type SBa-SBc in the present sample 54\% are of the {\it s} variety, 20\% are of type {\it rs} and 26\% are of the {\it r} variety. These numbers suggest that the fraction of all spiral galaxies that are of the {\it s}, {\it rs}, and {\it r} varieties does not differ significantly between normal (S) and barred (SB) field galaxies. 

For field spirals of types Sa to Sc the relative frequencies of the  
s and r varieties does not appear to depend significantly on parent 
galaxy luminosity. By the same token there is no evidence for a 
dependence of spiral arm variety on luminosity among barred spirals of 
types SBa to SBc. However, it is noted in passing that Ma (2002) has 
found that the pitch angle of spiral arms does depend on parent galaxy 
mass.

\section{ARM VARIETIES AND ENVIRONMENT}

Table 1 gives a contingency table that shows the frequencies of {\it s}
(spiral), {\it rs} (intermediate) and {\it r} (ring) type arms in ``field'',
``group'', and ``cluster'' environments for normal spirals of types 
Sa-Sc. From these data $\chi^{2}$ = 4.8 for four degrees of freedom, which    implies that there is no statistically significant relation between 
environment and arm variety. Similar information for the barred spirals 
of types SBa-SBc, that are contained in the northern Shapley-Ames 
catalog, are given in Table 2. From these data for barred spirals it 
is found that for $\chi^{2}$ = 1.9 for four degrees of freedom. In other 
words there appears to be no significant correlation between spiral 
arm variety and galaxy environment for either normal or for barred 
spirals. However, a caveat is that the frequency with which different 
arm varieties occur might itself be a function of Hubble type. Since 
the relative frequencies of Hubble types is a function of environment 
(Hubble 1936, Dressler 1980) it is therefore desirable to reinvestigate 
the possible dependence of spiral arm variety on environment by only 
using galaxies of a single Hubble type. For the present data set only
Sc galaxies are numerous enough for such a study. Table 3 and Table 4 
provide information on the possible dependence of spiral arm variety 
on environment for Sc and SBc galaxies, respectively. For Table 3 and 
Table 4 one finds $\chi^{2}$ values of 2.1 and 2.1 (with four degrees of 
freedom), respectively. This shows that there are no significant 
correlations between arm variety and environment for either Sc or for 
SBc galaxies.

\section{CONCLUSIONS}

The main result of the present investigation is that spiral arm 
variety does not appear to be a function of environment. In other words 
the internal properties of individual galaxies seem to determine
whether they will develop spiral structure of the {\it s}, {\it rs} or {\it r} varieties.   This conclusion is in complete agreement with the work of Buta (1999) who finds that ``[M]ost resonance rings are driven by internal nonaxisymmetric perturbations.''  According to Buta \& Combes (1996) most of such rings form by gas accumulation in resonances.  Among the northern Shapley-Ames galaxies classified by Sandage \& Tammann (1981), no correlations were found between galaxy environment 
and the arm varieties {\it s}, {\it rs}, and {\it r} for (1) all galaxies of types Sa-Sc,(2) all galaxies of types SBa-SBc, (3) galaxies of Hubble type Sc and (4) for galaxies of type SBc.

\clearpage

\begin{deluxetable}{crrr}
\tablecaption{Spiral arm variety versus galaxy environment for northern Shapley-Ames galaxies of Hubble types Sa-Sc}

\tablehead {\colhead{Arm Variety} & \colhead{Field} & \colhead{Group} & \colhead{Cluster}} 
\startdata

{\it s}       &       $ 114$    &    $38$     &     $75$  \\
{\it rs}      &       $  22$    &    $ 2$     &     $10$  \\
{\it r}       &       $  36$    &    $14$     &     $19$  \\

\enddata
\end{deluxetable}

\begin{deluxetable}{crrr}
\tablecaption{Spiral arm variety versus galaxy environment for northern Shapley-Ames galaxies of Hubble types SBa-SBc}

\tablehead{\colhead{Arm Variety} & \colhead{Field} & \colhead{Group} & \colhead{Cluster}} 
\startdata

{\it s}      &       $ 33$    &    $6$     &     $24$  \\
{\it rs}     &       $ 12$    &    $5$     &     $13$  \\
{\it r}      &       $ 16$    &    $3$     &     $11$  \\

\enddata
\end{deluxetable}

\begin{deluxetable}{crrr}
\tablecaption{Spiral arm variety versus galaxy environment for northern Shapley-Ames galaxies of Hubble type Sc}

\tablehead{\colhead{Arm Variety} & \colhead{Field} & \colhead{Group} & \colhead{Cluster}} 
\startdata

{\it s}       &       $ 71$    &    $25$    &     $46$  \\
{\it rs}      &       $ 12$    &    $1$     &     $5$  \\
{\it r}       &       $ 14$    &    $5$     &     $9$  \\

\enddata
\end{deluxetable}

\begin{deluxetable}{crrr}
\tablecaption{Spiral arm variety versus galaxy environment for northern Shapley-Ames galaxies of Hubble type SBc}

\tablehead{\colhead{Arm Variety} & \colhead{Field} & \colhead{Group} & \colhead{Cluster}} 
\startdata

{\it s}       &       $ 13$    &     $2$     &     $15$  \\
{\it rs}      &       $ 3$     &     $2$     &     $5$  \\
{\it r}        &       $ 3$     &     $1$     &     $2$  \\

\enddata
\end{deluxetable}

\end{document}